\documentclass{ws-procs975x65}

\usepackage{amsmath}
\usepackage{amsfonts}
\usepackage{amssymb}
\usepackage{epsfig} 
\usepackage{psfrag} 
\usepackage{color}

\begin{document}

\title{Aspects of Confinement in Low Dimensions}

\author{M.~J. BHASEEN AND A.~M. TSVELIK}

\address{Brookhaven National Laboratory, \\
Department of Physics, \\ 
Upton, NY 11973, USA\\ 
E-mail: bhaseen@bnl.gov, tsvelik@bnl.gov}

\address{
\vspace{0.5cm}
In Memory of Ian Kogan
\vspace{0.5cm}
\begin{figure}[h]
\begin{center}
\framebox{\epsfxsize=3.5cm\epsffile{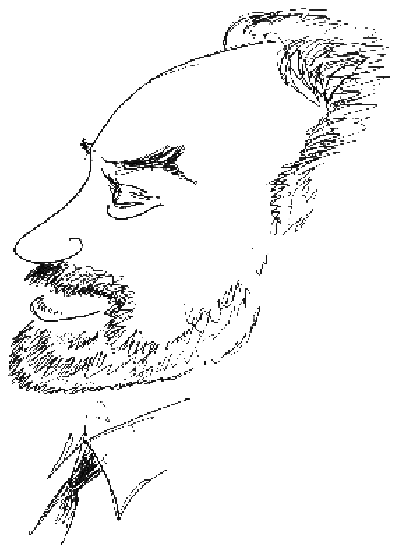}}
\end{center}
\end{figure}}


\maketitle

\abstracts{We briefly review some examples of confinement which arise in condensed matter physics. We focus on two instructive cases: the off-critical Ising model in a magnetic field, and an array of weakly coupled (extended) Hubbard chains in the Wigner crystal phase. In the appropriate regime, the elementary excitations in these 1+1 and quasi-one-dimensional systems are confined into `mesons'. Although the models are generically non-integrable, quantum mechanics and form factor techniques yield valuable information.}


\newpage

\section{Introduction}
The phenomenon of quark confinement lies at the center of modern particle physics. What is perhaps less widely known is that similar phenomena also emerge in condensed matter physics. In this brief review, we discuss some examples of `quark' confinement in 1+1 and quasi-one-dimensional systems. Celebrated cases include the Ising model in a magnetic field\cite{Mccoy:Breakup} and the spin-$1/2$ Heisenberg chain with frustration and dimerization\cite{Affleck:Confinement,Haldane:Spontaneous}. As is frequently the case in one-dimension, Lorentz invariance may emerge as an approximate symmetry of the low energy effective action. This opens the way to an assortment of field theory methods\cite{Tsvelik:QFT} and to a wealth of analytic results.

 The `quarks' in these, and  many other cases, appear as domain walls (or solitons) interpolating between different vacua. In particular, these excitations may carry `fractional' quantum numbers  which differ from those of the electron.  A prominent example are the excitations of the half-filled Hubbard model, where `spinons' carry spin-$1/2$ and no charge, and  `holons' carry charge $\pm e$ and no spin. The existence of these fractional excitations has been confirmed by numerous experiments on materials composed of weakly coupled chains, e.g. \cite{Zaliznyak:Spinons,YJune:RIXS}. 

Whilst these exotic excitations may be stabilized through the judicious choice of materials and parameter regime, they may form bound states in the presence of additional perturbations. In particular, the omnipresent interchain interactions may confine the elementary excitations and radically alter their masses and quantum numbers. A simple of illustration of this idea arises in an array of weakly coupled quantum Ising chains. The Hamiltonian governing a single chain is given by:
\begin{equation}
\label{quantumising}
H_{\parallel}=-J_{\parallel}\sum_{i} ({\hat\sigma}^z_{i}\hat\sigma^z_{i+1}+g\,\hat\sigma_{i}^x),
\end{equation}
where $J_\parallel>0$ is a ferromagnetic exchange constant, and the operators $\hat{\sigma}_i^{z,x}$  are Pauli matrices residing at site $i$. As is well known, this model undergoes a $T=0$ quantum phase transition at the point $g=1$\cite{Sachdev:QPT}. In particular, for $g<1$, the ground state degeneracy is spontaneously broken and the system develops long-range order $\langle\,\hat\sigma_i^z\,\rangle\neq 0$. Let us now consider the effect of a weak interaction between neighboring chains:
\begin{equation}
H_\perp=-J_\perp\sum_{ij}{\hat\sigma}_{i,j}^z\,{\hat\sigma}^z_{i,j+1}.
\end{equation}
Here, $J_\perp\ll J_\parallel$, and for notational simplicity we consider a two-dimensional array of chains. In the ordered regime, the effects of neighboring chains may be treated in a mean field approximation:
\begin{equation}
H_\perp^{{\rm MF}}=-\sum_{ij} h\,\sigma^z_{i,j}; \quad \quad h=\frac{1}{2}Z_\perp J_\perp\,\langle\,\sigma^z\,\rangle,
\end{equation}
where $Z_\perp$ is the transverse coordination number of the lattice. In this manner one obtains {\em decoupled} Ising chains in an {\em effective} magnetic field: 
\begin{equation}
H=\sum_i\left[-J_{\parallel}({\sigma}^z_i\sigma^z_{i+1}+g\,\sigma^x_i)-h\,\sigma_i^z\right].
\end{equation}
We emphasize that this is not an external magnetic field, which one may of course apply, but it arises quite naturally through the interchain interactions. As follows from the pioneering work of McCoy and Wu\cite{Mccoy:Breakup}, this weak magnetic field acts as a linearly confining potential on the zero field excitations:  the `quarks' are confined into a rich spectrum of `mesons'. Moreover, in this regime of weak confinement, analytic progress is possible.

The layout of this contribution is as follows. In \S \ref{IMF} we discuss the Ising model in a magnetic field in a little more detail. We focus on the properties in the ordered state where bound states form. In \S \ref{Q1DWC}
we discuss an example of a quasi-one-dimensional Wigner crystal: a weakly coupled array of quarter-filled extended Hubbard models. This system has much in common with both the Ising\cite{Mccoy:Breakup} and Heisenberg\cite{Affleck:Confinement} examples. In the limit of weak confinement, many `generations' of `mesons' are shown to exist. We conclude in \S \ref{CONC}.
\section{Ising Model in a Magnetic Field}
\label{IMF}
The Ising model has had a venerable history and there are excellent reviews devoted to it --- see for example\cite{McCoy:Book,Delfino:Ising}. In particular, the confining aspects of the 2D Ising model in a weak magnetic field were first exposed by McCoy and Wu\cite{Mccoy:Breakup}. The implications for gauge invariant correlation functions in the 2D ${\mathbb Z}_2$--gauge ${\mathbb Z}_2$--Higgs system were subsequently investigated\cite{McCoy:Gauge}. For a brief overview of these ideas see also page 106 of ref.~\cite{Mccoy:Connection}. 

In the ensuing discussion we contrast the behavior of the Ising model in the absence and presence of a magnetic field. In both cases we cast our discussion in terms of (suitably defined) form factors. The relevance of form factor techniques in this and other non-integrable models was first emphasized by Delfino, Mussardo and Simonetti\cite{Delfino:Nonintegrable}.


\subsection{$T\rightarrow T_c^-$ and $H=0$}
As is well known, in the scaling region close to criticality, the 2D Ising model is described by the field theory of a free Majorana fermion --- see for example \cite{Itzykson:Stat}: 
\begin{equation}
{\mathcal A}_{\rm FF}=\frac{1}{2\pi}\int d^2x \,\left[\Psi\bar\partial\Psi+\bar\Psi\partial\bar\Psi+im\bar\Psi\Psi\right].
\end{equation}
The fermion mass $m$ measures the departure from criticality, and we take $m>0$  in the `low-temperature' phase we are interested in.
The energy and momenta of these particles are conveniently parameterized in terms of the rapidity $\theta$
\begin{equation}
{\mathcal P}=m\sinh\theta, \quad {\mathcal E}=m\cosh\theta,
\end{equation}
where $\hbar=c=1$. In particular, they form an highly efficient basis in which to compute correlation functions\cite{Berg:Construction,Karowski:Exact,Karowshi:Physrep,Smirnov:Form}:
\begin{eqnarray}
\langle \,0\,|\,\sigma(\tau,x)\,\sigma(0,0)\,|\,0\,\rangle & = & \nonumber \\ 
& & \hspace{-3.5cm} \sum_{n=0,2,\dots}^\infty\int_{-\infty}^\infty \frac{d\theta_1\dots d\theta_n}{(2\pi)^n\,n!}\,\langle \,0\,|\,\sigma(\tau,x)\,|\,\theta_1\dots\theta_n\,\rangle \,\langle \,\theta_n\dots\theta_1\,|\,\sigma(0,0)\,|\,0\,\rangle.\label{ffapproach}
\end{eqnarray}
Here $(\tau,x)$ are Euclidean coordinates and $\sigma(\tau,x)$ is the continuum version of the lattice spin. In the ordered regime, the spin field only couples to intermediate states with an {\em even} number of particles.\footnote{The corresponding disorder operator $\mu(\tau,x)$ couples to an {\em odd} number of particles. By Kramers--Wannier duality the situation is reversed in the disordered `high-temperature' phase.} Equivalently,
\begin{eqnarray}
\label{expbelowtc}
\langle \,0\,|\,\sigma(\tau,x)\,\sigma(0,0)\,|\,0\,\rangle & = & \nonumber \\
& &  \hspace{-3cm} \sum_{n=0,2,\dots}^\infty\int_{-\infty}^\infty \frac{d\theta_1\dots d\theta_n}{(2\pi)^n\,n!}\,e^{iPx-E\tau}\,|\langle\,0\,|\,\sigma(0,0)\,|\,\theta_1\dots\theta_n\,\rangle|^2,
\end{eqnarray}
where 
\begin{equation}
P\equiv\sum_{i=1}^n m\,\sinh\theta_i, \quad E\equiv\sum_{i=1}^n m\,\cosh\theta_i. 
\end{equation}
The matrix elements of $\sigma(0,0)$ (or any other local field) between the vacuum and the multiparticle states, are known as form factors. The computation of form factors is central to all integrable models\cite{Berg:Construction,Karowski:Exact,Karowshi:Physrep,Smirnov:Form} and the lowest order contributions yield valuable information about the long distance correlations. In particular, the zero-particle form factor of the spin field is the spontaneous magnetization:
\begin{equation}
\label{isingvacex}
\langle \,0\,|\,\sigma(0,0)\,|\,0\,\rangle \equiv \langle\,\sigma\,\rangle=m^{1/8}\,\bar s,\quad {\bar s}=2^{1/12}e^{-1/8}A^{3/2},
\end{equation}
where $A=1.28243\dots$ is Glaisher's constant. Likewise the two-particle form factor is well known\cite{Berg:Construction}:
\begin{equation}
\label{isingtwopart}
\langle\,0\,|\,\sigma(0,0)\,|\,\theta_1\theta_2\,\rangle = i\,\langle\,\sigma\,\rangle\,\tanh\left(\frac{\theta_1-\theta_2}{2}\right).
\end{equation} 
The higher particle form factors are also known\cite{Berg:Construction} but they need not concern us here.\footnote{As in all integrable models, this involves solving the Form Factor Axioms\cite{Smirnov:Form} (or generalized Watson equations) with the appropriate S-matrix; in this case ${\rm S}=-1$.} Substituting (\ref{isingvacex}) and (\ref{isingtwopart}) into the expansion (\ref{expbelowtc}) one obtains
\begin{eqnarray}
\langle \,0\,|\,\sigma(\tau,x)\,\sigma(0,0)\,|\,0\,\rangle & = & \nonumber \\
& & \hspace{-3.5cm} |\langle\,\sigma\,\rangle|^2+|\langle\,\sigma\,\rangle|^2\int_{-\infty}^\infty\frac{2\,d\theta_+d\theta_-}{(2\pi)^22!}\,\tanh^2\theta_-\,e^{2m\,{\rm ch}\theta_-(ix\,{\rm sh}\,\theta_+-\tau\,{\rm ch}\,\theta_+)},
\end{eqnarray}
where $\theta_{\pm}=(\theta_1\pm\theta_2)/2$. Performing the integral over $\theta_+$,
\begin{equation}
\label{ising02exp}
\langle \,0\,|\,\sigma(\tau,x)\,\sigma(0,0)\,|\,0\,\rangle=|\langle\,\sigma\,\rangle|^2+\frac{|\langle\,\sigma\,\rangle|^2}{2\pi^2}\,\int_{-\infty}^\infty d\theta\,\tanh^2\theta\,{\rm K}_0(2\bar r\,{\rm ch}\,\theta),
\end{equation}
where $\bar r\equiv mr$ and $r\equiv \sqrt{x^2+\tau^2}$.
The integral in (\ref{ising02exp}) may be evaluated in closed form. This yields the famous result of Wu and collaborators\cite{Wu:Spin}:
\begin{equation}
\label{isingexplicitexp}
\langle \,0\,|\,\sigma(\tau,x)\,\sigma(0,0)\,|\,0\,\rangle = |\langle\,\sigma\,\rangle|^2\,{\mathcal G}(\bar r),
\end{equation}
where the scaling function
\begin{equation}
{\mathcal G}(\bar r)=1+ \frac{1}{\pi^2}\left({\bar r}^2\left[{\rm K}_1^2(\bar r)-{\rm K}_0^2(\bar r)\right]-{\bar r}{\rm K}_0(\bar r)K_1(\bar r)+\frac{1}{2}{\rm K}_0^2(\bar r)\right)+\dots
\end{equation}
bears a remarkable connection with the solution of the Painlev{\'e} III equation \cite{Barouch:Zerofield,Wu:Spin}.
The large distance asymptotics of the spin-spin correlation function are readily extracted:
\begin{equation}
\label{isingasymp}
\langle \,0\,|\,\sigma(\tau,x)\,\sigma(0,0)\,|\,0\,\rangle\sim |\langle\,\sigma\,\rangle|^2\left\{1+\frac{1}{8\pi}\frac{e^{-2\bar r}}{{\bar r}^2}+{\mathcal O}(e^{-4\bar r})\right\}; \quad \bar r\rightarrow\infty.
\end{equation}
This example helps convey the efficiency of the form factor approach.

\begin{figure}[ht]
\centerline{\epsfxsize=8cm\epsffile[48 21 738 560]{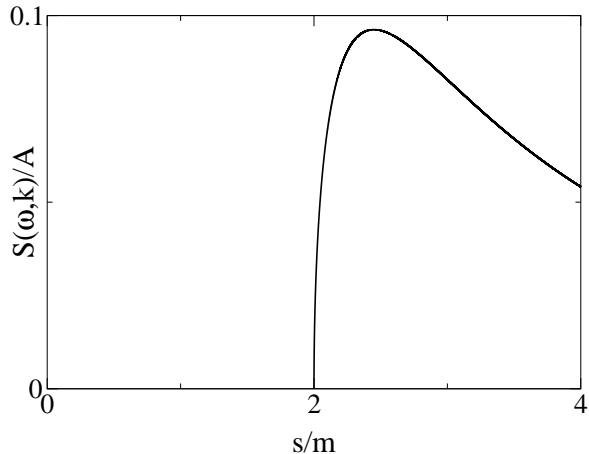}}   
\caption{Dynamical structure factor of the Ising model in the ordered regime and in the absence of a magnetic field; $s\equiv\sqrt{\omega^2-k^2}$ and $A\equiv |\langle\,\sigma\,\rangle|^2/m^2$. 
\label{isingzf}}
\end{figure}

In condensed matter applications it is useful to introduce the so-called {\em dynamical susceptibility}
\begin{equation}
\label{defchi}
\chi(\omega,k)=\chi_E(\bar\omega,k)_{\bar\omega\rightarrow\varepsilon-i\omega},
\end{equation}
where $\chi_E(\bar\omega,k)$ is nothing but the Fourier transform of the Euclidean spin-spin correlation function
\begin{equation}
\chi_E(\bar\omega,k)=-\int_{-\infty}^{\infty}dx\int_{-\infty}^{\infty} d\tau\,e^{i\bar\omega\tau-ikx}\,\langle{\rm T}_{\tau}\,\sigma(\tau,x)\sigma(0,0)\,\rangle.
\end{equation}
${\rm T}_\tau$ denotes time ordering. The definition (\ref{defchi}) includes the analytic continuation to real frequencies and $\varepsilon$ is a positive infinitesimal. The {\em dynamical structure factor}, as measured by inelastic neutron scattering, is extracted from this:
\begin{equation}
S(\omega,k)\equiv -{\rm Im}\,\chi(\omega,k).
\end{equation}
In this way, the two-particle contribution to the dynamical structure factor may be obtained
\begin{equation}
S(\omega,k)=\,|\langle\,\sigma\,\rangle|^2\,\frac{\sqrt{\omega^2-k^2-4m^2}}{(\omega^2-k^2)^{3/2}},
\end{equation}
where $\omega>0$. We plot this in Fig.~\ref{isingzf}. 

As is well known, the dynamical structure factor is a direct reflection of the elementary excitations. Indeed, the absence of any sharp peaks in Fig.~\ref{isingzf} reveals that they are {\em deconfined} -- the spin operator couples to {\em pairs} of excitations and a two-particle {\em continuum} exists above threshold.

In the next section we recall the dramatic change of this picture in the presence of a weak magnetic field\cite{Mccoy:Breakup}. 
\subsection{$T\rightarrow T_c^-$ and $H\gtrsim 0$}
In the presence of a magnetic field the scaling region may be described by the non-integrable\footnote{At $T=T_c$ the model is actually integrable and yields the celebrated ${\rm E}_8$ mass spectrum\cite{Zamolodchikov:Integrals,Fateev:Simplylaced}.} Ising Field Theory:
\begin{equation}
\label{IFT}
{\mathcal A}_{\rm IFT}={\mathcal A}_{{\rm FF}}+\hat h\int d^2x\,\sigma(x),
\end{equation}
where ${\mathcal A}_{\rm FF}$ is the action of free massive fermions. 
As follows from McCoy and Wu\cite{Mccoy:Breakup}, the effect of the magnetic field in (\ref{IFT}) is to confine the massive free fermions or `quarks' into bound states or `mesons'. In the limit of small magnetic fields, the masses of these bound states follow from the Schr\"odinger equation of two particles of mass $m$ subject to a linear confining potential:
\begin{equation}
\label{linschrod}
-\frac{1}{m}\frac{d^2\Psi_b(x)}{dx^2}+\lambda|x|\Psi(x)={\mathcal E}_b\Psi_b(x),\end{equation}
where we employ the reduced mass. In this example, the confining parameter $\lambda$ or `string tension' is related to the magnetic field by \cite{Mccoy:Breakup,Fonseca:Ising}
\begin{equation}
\label{lambdahrel}
\lambda=2\hat h\,\langle \,\sigma\,\rangle,
\end{equation}
and the corresponding bound states have masses given by
\begin{equation}
m_b=2m+{\mathcal E}_b.
\end{equation}
As follows from (\ref{IFT}) $\lambda$ has dimensions of $[{\rm mass}]^2$ ($\hbar=c=1$) and equation (\ref{linschrod}) is consistent on dimensional grounds. In accordance with the more recent work of Fonseca and Zamolodchikov \cite{Fonseca:Ising}, equations (\ref{linschrod}) and (\ref{lambdahrel}) may be derived from the action of (\ref{IFT}) on an appropriate two-particle bound state. Their method is rather general and employs the known finite size form factors of the perturbing field ($\sigma$) in the unperturbed (free fermion) model. In particular it permits a systematic study of the relativistic Bethe--Salpeter corrections to equation (\ref{linschrod}) and the resulting bound state mass spectrum. That is to say, the Schr\"odinger equation (\ref{linschrod}) is a small momentum approximation which holds for small magnetic fields or weak confinement. For our present discussion however, equations (\ref{linschrod}) and (\ref{lambdahrel}) are sufficient.

The Schr\"odinger equation (\ref{linschrod}) is easily solved \cite{Landau:QM}. In the region $x\ge 0$ we introduce the change of variables 
\begin{equation}
\xi=(m\lambda)^{1/3}(x-{\mathcal E}/\lambda)
\end{equation}
so as to yield the Airy equation
\begin{equation}
\frac{d^2\Psi(\xi)}{d\xi^2}-\xi\Psi(\xi)=0.
\end{equation}
The normalizable solutions may be written in terms of an Airy function \cite{Abramowitz:Tables}:
\begin{equation}
\label{airywavef}
\Psi(\xi)\propto \,{\rm Ai}(\xi).
\end{equation}
As expected, this wavefunction  is oscillatory in the classically allowed region $x<{\mathcal E}/\lambda$ and damped in the classically forbidden region $x>{\mathcal E}/\lambda$.
Since the potential $\lambda|x|$ is an even function, the wavefunctions have a definite symmetry and must be matched smoothly at the origin. In particular, the antisymmetric wavefunctions must vanish at $x=0$. This leads to the energy level quantization condition:
\begin{equation}
{\rm Ai}\left[-\left(\frac{m}{\lambda^2}\right)^{1/3}{\mathcal E}_b\right]=0.
\end{equation}
 The bound states of the Ising model in a weak magnetic field are therefore indexed by the zeroes of the Airy function \cite{Mccoy:Breakup}. Their masses are given by
\begin{equation}
\label{airymass}
m_b=2m+\left(\frac{\lambda^2}{m}\right)^{1/3}Z_b;\quad {\rm Ai}(-Z_b)=0.
\end{equation}
Equivalently, in notations closer to those of McCoy and Wu\cite{Mccoy:Breakup}
\begin{equation}
\label{mccoymass}
m_b=2m+\frac{(\hat h\,\langle\,\sigma\,\rangle\Lambda_b)^{2/3}}{m^{1/3}}; \quad J_{\frac{1}{3}}\left(\tfrac{1}{3}\Lambda_b\right)+J_{-\frac{1}{3}}(\tfrac{1}{3}\Lambda_b)=0,\end{equation}
where we have used (\ref{lambdahrel}) and $Z_b\equiv (\Lambda_b/2)^{2/3}$. The absence of the symmetric wavefunctions from the known Ising model mass spectrum (\ref{mccoymass}) is apparent, and we shall return to this point in the next section. As $\lambda\rightarrow 0$ there is a proliferation in the number of bound states and their masses densely populate the interval between $2m$ and $4m$; in general, when the mass of a bound state exceeds twice the mass of the lightest `meson' $2m_1$ they become unstable. As $\lambda\rightarrow 0$ we confine our attention to bound states in the vicinity of threshold $2m$.
 
\begin{figure}[ht]
\centerline{\epsfxsize=10cm\epsfbox{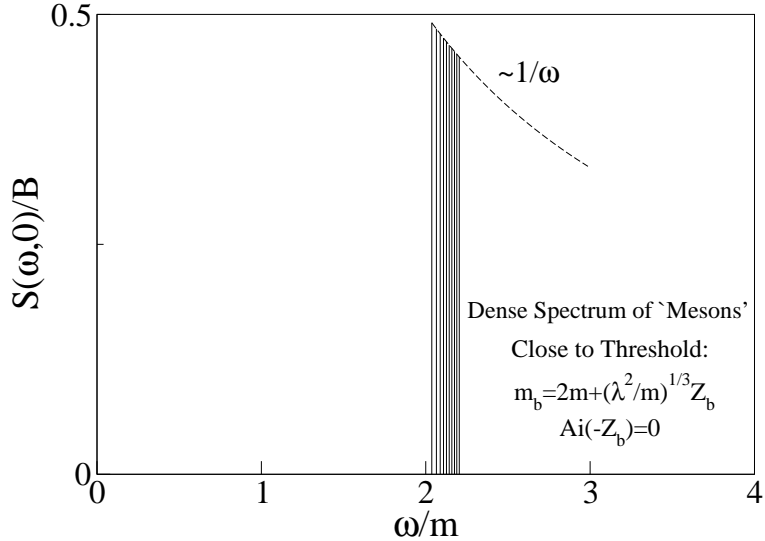}}   
\caption{Ising model dynamical structure factor in the ordered regime and in a weak magnetic field. For illustration we have replaced the Dirac delta functions by unity. $B\equiv\pi\bar h|\langle\,\sigma\,\rangle|^2/m$ and we have set $\bar h=0.001$.
\label{isingsffield}}
\end{figure}

Having discussed the spectrum of the Ising model in a weak magnetic field, we now turn our attention to the spin-spin correlation function. The confining magnetic field yields a spectrum of bound states of mass $m_b$. On general grounds we expect the spin-spin correlation function to have the form:
\begin{eqnarray}
\langle\,0\,|\,\sigma(\tau,x)\,\sigma(0,0)\,|\,0\,\rangle_\lambda & = & \nonumber \\
& & \hspace{-3.5cm} |\langle\,\sigma\,\rangle|^2 +\sum_{b=1}^{n_b}\int_{-\infty}^\infty\frac{d\theta}{2\pi}\,e^{im_b\,{\rm sh}\,\theta \,x-m_b\,{\rm ch}\,\theta\,\tau}\,|\langle\,0\,|\,\sigma(0,0)\,|\,\Psi_b(\theta)\,\rangle|^2,
\end{eqnarray}
where $|\Psi_b(\theta)\rangle$ is an asymptotic state (to be discussed in the next section) describing the non-trivial bound state. For a general operator ${\mathcal O}$ of spin-$s$, we expect
\begin{equation}
\label{ffspins}
\langle\,0\,|\,{\mathcal O}\,|\,\Psi_b(\theta)\,\rangle = \langle\,0\,|\,{\mathcal O}\,|\,\Psi_b\,\rangle\,e^{s\theta}.
\end{equation}
Since $\sigma$ is a spinless operator, the matrix elements (\ref{ffspins}) are independent of rapidity.
Performing the integral over rapidity yields:
\begin{equation}
\label{spinspinconf}
\langle\,0\,|\,\sigma(\tau,x)\,\sigma(0,0)\,|\,0\,\rangle_\lambda=|\langle\,\sigma\,\rangle|^2+\frac{1}{\pi}\sum_{b=1}^{n_b}|\langle\,0\,|\,\sigma(0,0)\,|\,\Psi_b \rangle|^2\,{\rm K}_0(m_br);
\end{equation}
where $n_b$ is the number of stable bound states. The main problem of course is that we do not know the rapidity independent matrix elements appearing in (\ref{spinspinconf}), let alone their multiparticle extensions. After all, the Ising field theory (\ref{IFT}) is non-integrable, and the powerful axiomatic approach to integrable models\cite{Smirnov:Form} does not apply. It is however a perturbation of a well understood integrable model. In the next section we shall illustrate how such single particle bound state matrix elements may be calculated directly in the limit of weak confinement. Understanding their detailed form is an obvious task in a more systematic approach to non-integrable models with confinement. Before embarking on this pursuit however, let us proceed a little with the benefit of hindsight. In particular, let us focus on the long distance asymptotics of (\ref{spinspinconf}). We know that as the confinement parameter $\lambda\rightarrow 0$ we must recover the asymptotics (\ref{isingasymp}) of the unperturbed system. Indeed, this important cross-check was performed in the seminal work of McCoy and Wu \cite{Mccoy:Breakup}. In addition, let us assume that as $\lambda\rightarrow 0$ {\em the matrix elements are the same for all bound states}; we shall justify this below. In this manner, the long distance asymptotics of the {\em connected} contribution to (\ref{spinspinconf}) read:
\begin{equation}
\langle\,0\,|\,\sigma(\tau,x)\,\sigma(0,0)\,|\,0\,\rangle_{\lambda\rightarrow 0}^c \sim \frac{|\langle\,0\,|\,\sigma(0,0)\,|\,\Psi_b\,\rangle|^2}{\sqrt{2\pi}} \sum_{b=1}^{n_b}\frac{e^{-m_br}}{\sqrt{m_br}}\label{asymsum}.
\end{equation}
As $\lambda\rightarrow 0$, the bound state masses (\ref{airymass}) become closer together. Put another way, the bound state poles in momentum space must coalesce to form a branch cut \cite{Mccoy:Breakup}. In this limit one may therefore convert the sum in (\ref{asymsum}) to an integral:
\begin{equation}
\label{bsffintrep}
\sum_{b=1}^{n_b}\frac{e^{-m_br}}{\sqrt{m_br}}\rightarrow  \frac{e^{-2mr}}{\sqrt{2mr}}\int_0^\infty db \,\exp\left[{-r\left(\frac{3\pi\lambda\, b}{2\sqrt{m}}\right)^{2/3}}\right],
\end{equation}
where we have used the fact that the zeroes of the Airy function have the limiting form \cite{Abramowitz:Tables}
\begin{equation}
Z_b\approx\left[3\pi/2\left(b-1/4\right)\right]^{2/3}; \quad b\rightarrow\infty.\end{equation}
This yields the long distance behavior
\begin{equation}
\langle\,0\,|\,\sigma(\tau,x)\,\sigma(0,0)\,|\,0\,\rangle_{\lambda\rightarrow 0}^c\sim \frac{|\langle\,0\,|\,\sigma(0,0)\,|\,\Psi_b\,\rangle|^2}{4\pi\lambda}\frac{e^{-2mr}}{r^2}.
\end{equation}
In order to recover the correct asymptotics (\ref{isingasymp}) as the confining perturbation is turned off, we {\it require} that the bound state matrix elements have the specific form: 
\begin{equation}
\label{nonintamp}
|\langle\,0\,|\,\sigma(0,0)\,|\,\Psi_b\,\rangle| =\frac{\langle\,\sigma\,\rangle}{m}\,\sqrt{\frac{\lambda}{2}};\quad \lambda\rightarrow 0.\end{equation}
Equivalently,
\begin{equation}
\label{hmatrixel}
|\langle\,0\,|\,\sigma(0,0)\,|\,\Psi_b\,\rangle| =\langle\,\sigma\,\rangle\sqrt{\bar h};\quad \hat h\rightarrow 0.
\end{equation}
where $\bar h=\langle\,\sigma\,\rangle \hat h/m^2$.
Indeed, substituting (\ref{hmatrixel}) into (\ref{asymsum}) we recover the expected asymptotics of McCoy and Wu\cite{Mccoy:Breakup}: 
\begin{equation}
\langle\,0\,|\,\sigma(\tau,x)\,\sigma(0,0)\,|\,0\,\rangle_{\hat h}^c\sim \bar h\,|\langle\,\sigma\,\rangle|^2 \,\frac{e^{-2\bar r}}{2\sqrt{\pi\bar r}}\sum_b e^{-(\bar h\Lambda_b)^{2/3}{\bar r}}.
\end{equation}
It is readily verified that the weak field matrix elements (\ref{hmatrixel}) lead to the dynamical structure factor:
\begin{equation}
S(\omega,k)=\pi\,\bar h\,|\langle\,\sigma\,\rangle|^2\,\sum_b\frac{\delta\left(\omega-\sqrt{m_b^2+k^2}\right)}{\omega}.
\end{equation}
We illustrate this dependence in Fig.~\ref{isingsffield}. The two-particle continuum of Fig.~\ref{isingzf} has been replaced by a series of sharp peaks --- the elementary excitations are now {\em confined}.

In the next section we shall continue our discussion of the bound state matrix elements (\ref{nonintamp}) and (\ref{hmatrixel}). We shall demonstrate how they (and indeed those in other models) follow from a direct computation within the so-called `two-quark' approximation\cite{Fonseca:Ising}. In the same spirit as the relativistic mass corrections have been studied by Fonseca and Zamolodchikov \cite{Fonseca:Ising} matrix element corrections may in principle be obtained. Their effect on the long distance asymptotics is presumably less significant however, and leading order results may be sufficient for many purposes.
\subsection{Bound State Amplitudes}
\label{BSA}
As we see from (\ref{ffspins}) the amplitude we seek is determined by the overlap between the vacuum, the operator, and the bound state in its {\em rest frame}:\begin{equation}
\langle\,0\,|\,{\mathcal O}\,|\,\Psi_b\,\rangle.
\end{equation}
The main problem is that we do not have an exact handle on the bound state $|\,\Psi_b\,\rangle$. However, following Fonseca and Zamolodchikov\cite{Fonseca:Ising} for weak confinement (and ${\mathcal E}_b\equiv m_b-2m\ll m$) we may consider the `two-quark' approximation:
\begin{equation}
\label{twoquark}
|\,\Psi_b\,\rangle=\frac{1}{\sqrt{m}}\int_{-\infty}^\infty\frac{dq}{2\pi}\,\Psi_b(q)\,Z^\dagger(q)Z^\dagger(-q)\,|\,0\,\rangle,
\end{equation}
where $Z^\dagger(q)$ are the Faddeev--Zamolodchikov creation operators of the constituent `quarks'. In the case at hand these are fermionic creation operators, but in general they furnish rapidity dependent commutation relations. They may also carry isotopic indices. The function $\Psi_b(q)$ is the Fourier transform of the  normalized wavefunction (\ref{airywavef}): 
\begin{equation}
\Psi_b(q)=\int_{-\infty}^{\infty}dx\,e^{iqx}\,\Psi_b(x),\quad {\rm where} \quad \int_{-\infty}^{\infty} dx\,|\Psi_b(x)|^2=1.
\end{equation}
In our conventions, the overall factor of $1/\sqrt{m}$ is required on dimensional grounds. An illuminating application of such a wavefunction to bound state mass spectra may be found in the work Gat and Rosenstein\cite{Gat:New}; the integrable 1+1 dimensional massive Thirring model is viewed as a problem with delta-function confinement and the bound state meson is nothing but the sine-Gordon particle. In the present context, the approximation (\ref{twoquark}) has the virtue that it embodies both the integrable aspects of the unperturbed system together with the most important effect of (non-integrable) confinement.  As follows from the `two-quark' approximation, the matrix element of the spin field is given by:\begin{equation}
\label{spinel}
\langle\,0\,|\,\sigma(0,0)\,|\,\Psi_b\,\rangle=\frac{1}{\sqrt{m}}\int_{-\infty}^{\infty} dx\,\Psi_b(x)\int_{-\infty}^\infty\frac{dq}{2\pi}\,e^{iqx}\,\langle\,0\,|\,\sigma(0,0)\,|\,q,-q\,\rangle,
\end{equation}
where the normalized antisymmetric wavefunctions take the form: 
\begin{equation}
\label{normalized}
\Psi_b(x)=\frac{{\rm Ai}(\xi_b)}{\sqrt{2}\,(m\lambda)^{1/6}|{\rm Ai}^\prime(-Z_b)|};\quad x>0.
\end{equation}
Substituting $q=m\sinh\theta$ into the Ising form factor (\ref{isingtwopart}) one obtains
\begin{equation}
\langle\,0\,|\,\sigma(0,0)\,|\,q,-q\,\rangle=i\,\langle\,\sigma\,\rangle\frac{q}{\sqrt{m^2+q^2}}.
\end{equation}
The Schr\"odinger description is valid for small momenta\cite{Fonseca:Ising} and in view of this we expand the form factor in powers of momentum:
\begin{equation}
\label{ffexpand}
\langle\,0\,|\,\sigma(0,0)\,|\,q,-q\,\rangle=i\,\langle\,\sigma\,\rangle\,\frac{q}{m}+{\mathcal O}\left(\frac{q}{m}\right)^3.
\end{equation}
To lowest order equation (\ref{spinel}) becomes:
\begin{equation}
\label{ampderiv}
\langle\,0\,|\,\sigma(0,0)\,|\,\Psi_b\,\rangle=-\frac{\langle\,\sigma\,\rangle}{m^{\frac{3}{2}}}\left.\frac{d\Psi_b(x)}{dx}\right|_{x=0}.
\end{equation}
Substituting in the explicit form of the wavefunction (\ref{normalized}) yields the bound state matrix element:
\begin{equation}
\label{ampoverlap}
\langle\,0\,|\,\sigma(0,0)\,|\,\Psi_b\,\rangle=(-1)^b\,\frac{\langle\,\sigma\,\rangle}{m}\,\sqrt{\frac{\lambda}{2}},
\end{equation}
in agreement with equation (\ref{nonintamp}). There are a couple of things to note in this relatively simple example. First, to lowest order the bound state dependence has disappeared: the normalization of the Airy wavefunction (\ref{normalized}) has canceled under the differential action of the form factor (\ref{ffexpand}). Second, the derivative of a {\em symmetric} wavefunction must vanish at the origin, and by equation (\ref{ampderiv}) would yield a zero matrix element. 

We note that whilst 
${\mathcal O}({\mathcal E}_b/m)$ corrections to (\ref{ampoverlap}) are easily obtained from the expansion (\ref{ffexpand}), an accurate determination of the coefficients presumably requires the Bethe--Salpeter approach of Fonseca and Zamolodchikov\cite{Fonseca:Ising}. This lies beyond the scope of this current work. This therefore completes our brief discussion of the Ising model in a magnetic field. We turn our attention to another model of condensed matter physics which also exhibits aspects of confinement.

\section{Quasi-1D Wigner Crystals}
\label{Q1DWC}
 In this section we describe a possible realization of confinement in Wigner crystals\cite{Wigner:Interact}.  If the average potential energy of an electron system exceeds its kinetic energy, the electrons themselves may adopt a regular arrangement in space and thus form a crystal.  Wigner (or electron) crystallization is a spectacular consequence of electron--electron interactions. 

One may gain some insight into Wigner crystallization from a simple one-dimensional example --- the extended Hubbard model at quarter-filling:
\begin{eqnarray}
\label{exthub}
H & = & -t\sum_{i,\sigma}(c_{i,\sigma}^\dagger c_{i+1,\sigma}+{{\rm h}.{\rm c}.})+U\sum_i n_{i,\uparrow}\,n_{i,\downarrow} \nonumber \\  & & \hspace{0.6cm} + V_\parallel\sum_{i,\sigma,\sigma^\prime}\left(n_{i,\sigma}-1/4\right)\left(n_{i+1,\sigma^\prime}-1/4\right).
\end{eqnarray}
$c^\dagger_{i,\sigma}$ is an electron creation operator and $n_{i,\sigma}\equiv c^\dagger_{i,\sigma} c_{i,\sigma}$ is the number operator. The electrons move on a rigid\footnote{For simplicity we neglect the possibility of lattice distortions or phonons.} lattice with sites labeled by $i$. They carry spin $\sigma=\uparrow,\downarrow$ and obey the usual anticommutation relations. The most important effects of the long-range 1D Coulomb interaction are described by the on-site repulsion $U$ and the nearest neighbor ``extended'' repulsion $V_\parallel$. The model is defined to have on average one electron for every two sites; in view of the Pauli principle, the band is quarter-filled and $k_F=\pi/4$. 
For related works see for example\cite{Mila:Phase,Penc:Phase,Giamarchi:Mott,Nakamura:Tricritical,Essler:Quarter,Essler:Spectral,Sano:Combined}

For small values of $U$ and $V_\parallel$ the system (\ref{exthub}) is a metal. Indeed, at quarter-filling, elementary band theory predicts metallic behavior. However, as the interactions are increased, this system exhibits a metal-insulator transition\cite{Ovchinikov:2t,Mila:Phase,Penc:Phase}. This $T=0$ {\em quantum phase transition}\cite{Sachdev:QPT} is driven by $8k_F$ Umklapp processes. Such processes become relevant only in the presence of sufficiently strong repulsion. The insulating state, which arises due to interactions rather than band filling, is an example of a {\em Mott insulator}\cite{Mott:MIT}.

A glimpse into the nature of the Mott transition and the associated Wigner crystal, is possible in the limit $U\rightarrow\infty$ \cite{Ovchinikov:2t}. 
In so far as the charge degrees of freedom are concerned, the model with $V_\parallel=0$ may be mapped onto a model of {\em free spinless fermions}. Including the nearest neighbor repulsion one obtains a one-dimensional model of {\em interacting spinless fermions}\cite{Ovchinikov:2t}:
\begin{equation}
\label{spinlessfermion}
H_c=-t\sum_{i}(c_i^\dagger c_{i+1}+{\rm h.c.})+V_{\parallel}\sum_i \left(n_i-1/2\right)\left(n_{i+1}-1/2\right),
\end{equation}
where $n_i\equiv\sum_{\sigma}n_{i,\sigma}$. As is well known \cite{Tsvelik:QFT,Sachdev:QPT} this model maps on to the spin-$1/2$ ${\rm XXZ}$ spin chain via the Jordan--Wigner  transformation:
\begin{equation}
\label{XXZ}
H^{{\rm XXZ}}_c =J\sum_i(S^x_iS^x_{i+1}+S^y_iS^y_{i+1}+\Delta\,S^z_iS^z_{i+1}). 
\end{equation}
The parameters are related by
\begin{equation}
J=-2t,\quad  \Delta=-\frac{V_\parallel}{2t}. 
\end{equation}
With {\em repulsive} fermions both $J$ and $\Delta$ are {\em negative}. With this sign of $J$, it is known that for $-1<\Delta<0$ the chain (\ref{XXZ}) is gapless, whereas for $\Delta<-1$ it develops both a gap and long-range antiferromagnetic Ising order:
\begin{equation}
\label{XXZlro}
\langle\,(-1)^i S^z_i\,\rangle \neq 0.
\end{equation}

\begin{figure}[ht]
\centerline{\epsfxsize=10cm\epsffile[28 15 718 560]{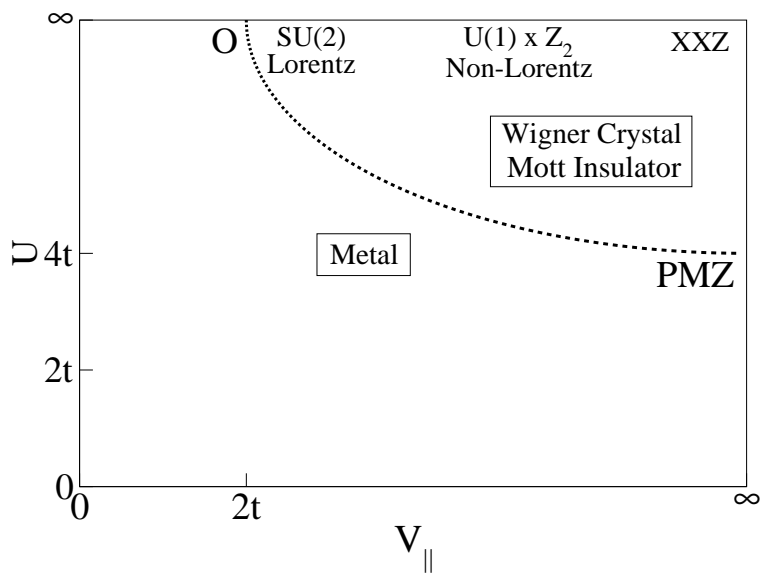}}  
\caption{\label{fig:wigner} Metal-Insulator transition in the 1D extended Hubbard model and the associated Mott insulating Wigner crystal or $4k_F$ CDW state. The phase boundary connects the ``Ovchinikov point'' $(O)$ to the ``Penc--Mila--Zotos point'' (PMZ). Along the ``XXZ line'' $(U\rightarrow\infty)$ the system undergoes a BKT transition at the point $O$ .}
\end{figure}

 The gap implies that the spinless fermion model (\ref{spinlessfermion}) is a {\em Mott insulator} for $V_\parallel >2t$, and a {\em metal} for $V_\parallel< 2t$. That is to say, as $U\rightarrow\infty$ in our original model (\ref{exthub}) the Mott transition occurs for $V_\parallel=2t$\cite{Ovchinikov:2t}. This is confirmed by numerical simulation\cite{Penc:Phase}. As follows from the mapping to the XXZ chain, this metal-insulator transition is of the Berezinskii--Kosterlitz--Thouless (BKT) type. It also follows that close to the transition (when the insulating gap is much smaller than the bandwidth) the model is ${\rm SU}(2)$ invariant. At low energies we may also ``linearize the fermionic spectrum'' \cite{Tsvelik:QFT,Tsvelik:Boson} and $1+1$-dimensional Lorentz invariance emerges. Away from the transition the ${\rm SU}(2)$ symmetry is broken down to ${\rm U}(1)\times{\mathbb Z}_2$, and the Lorentz invariant limit is no longer justified. We note that as $U$ is lowered, one expects this critical value of $V_\parallel$ to increase. Indeed, the numerical simulations indicate that the phase boundary connects the ``Ovchinikov point'' $(U,V_\parallel)=(\infty,2t)$ and the ``Penc--Mila--Zotos point'' $(U,V_\parallel)=(4t,\infty)$\cite{Mila:Phase,Penc:Phase} --- see Fig.~\ref{fig:wigner}. It is plausible that model (\ref{spinlessfermion}) may also describe the transition for finite $U>4t$, providing $V_\parallel$ is replaced by an effective repulsion $V_\parallel^{\rm eff}(U)$. 
 
\begin{figure}[ht]
\centerline{\epsfxsize=7cm\epsffile[0 0 460 190]{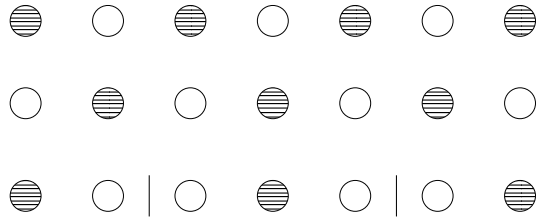}}  
\caption{\label{fig:kogancdw} Degenerate Wigner crystal (or CDW) ground states in the Mott insulating regime. The darker regions are sites of higher electron density. The elementary excitations, or `quarks', are domain walls separating these ground states.}
\end{figure}

Since $S^z_i=n_i-1/2$, the ordering (\ref{XXZlro}) of the {\rm XXZ} chain implies that the density of spinless fermions (\ref{spinlessfermion}) alternates from site to site (see Fig.~\ref{fig:kogancdw}) and 
\begin{equation}
\label{chargeexp}
\langle\,(-1)^in_i\,\rangle\neq 0.
\end{equation}
i.e. the insulator is a Wigner crystal or ``$4k_F$  CDW'' (Charge Density Wave). Indeed, the interaction term in (\ref{spinlessfermion}) clearly favors this alternation. As shown in Fig.~\ref{fig:kogancdw} the ground state is doubly degenerate with respect to charge ordering. 
The elementary excitations or `quarks' in this insulating regime  are domain walls separating these two ground states. It may be seen that these `quarks' carry fractional charge $\pm e/2$.
Moreover, it may be shown that these `quarks' are {\em deconfined}. As in the Ising model, this manifests itself in the absence of poles in the appropriate dynamical (charge) response functions. In the case at hand these are the spectral function\cite{Essler:Quarter,Essler:Spectral} or the optical conductivity. The relevant probes couple to {\em pairs} of `quarks'.

This concludes our discussion of the purely one-dimensional system (\ref{exthub}). In the next section we shall discuss what happens when weak interchain interactions are switched on.
\subsection{Interchain Interactions}
Let us take our model of spinless fermions (\ref{spinlessfermion}) and switch on {\em weak} interchain interactions, $V_\perp\ll V_\parallel$:
\begin{eqnarray}
H_c & = & -t\sum_{ij}(c^\dagger_{i,j}\,c_{i+1,j}+{\rm h.c.}) + V_\parallel \sum_{ij} (n_{i,j}-1/2)(n_{i+1,j}-1/2) \nonumber \\
& & \hspace{2cm} + V_\perp\sum_{ij}(n_{i,j}-1/2)(n_{i,j+1}-1/2).
\end{eqnarray}
For notational simplicity we consider a two-dimensional array of chains. The extension to three-dimensions is straightforward. We also neglect the possibility of interchain hopping. In the ordered phase, we may treat the interchain interactions in a mean field approximation. That is to say, from the perspective of a single chain, we replace the neighboring chains by a suitable expectation value. This yields a set of {\em decoupled} chains in a staggered chemical potential:
\begin{eqnarray}
\label{qfhmf}
H_c & = & -t\sum_{i}(c_i^\dagger c_{i+1}+c_{i+1}^\dagger c_{i}) \nonumber \\
& & +V_{\parallel}\sum_i \left(n_i-1/2\right)\left(n_{i+1}-1/2\right)+g_\perp\sum_i(-1)^i\,n_i,
\end{eqnarray}
where
\begin{equation}
\label{gperp}
g_\perp=\frac{1}{2}\,Z_\perp V_\perp\langle \,(-1)^i\,n_i\,\rangle,
\end{equation}
and $Z_\perp$ is the transverse coordination number. As follows from $(\ref{chargeexp})$ the expectation value occurring in (\ref{gperp}) is non-zero even for $V_\perp\rightarrow 0$. This interaction renders the metal-insulator transition first order. 

We shall discuss the model (\ref{qfhmf}) in the limit where the dynamically generated `quark' mass is much smaller than the bandwidth, $m\ll t$. In this limit, close to the transition, Lorentz invariance emerges and field theory may be applied. In the vicinity of the ${\rm SU}(2)$ invariant Ovchinikov point the mean field Hamiltonian (\ref{qfhmf}) may be bosonized\cite{Tsvelik:Boson}. The corresponding mean field\footnote{The most essential corrections to the mean field action couple neighboring chains $n$ and $m$ via an interaction $\sim V_\perp\int d^2x\,\,\partial_x\Phi_c^n\partial_x\Phi_c^{m}$. This lifts the ${\rm SU}(2)$ degeneracy and also leads to an anisotropy in the optical conductivity tensor $\sigma^{xx}\neq\sigma^{yy}$ .} action may be written:
\begin{equation}
\label{dsg}
{\mathcal A}_{c}^{\rm DSG}=\int d^2x\,\left[\frac{1}{16\pi}(\partial_\mu\Phi_c)^2-\hat\mu\,\cos\beta\,\Phi_c+\hat\lambda\,\cos\frac{\beta}{2}\,\Phi_c\right],
\end{equation}
where in this normalization $\beta^2=1$. This model is the well studied one-dimensional {\em double sine-Gordon model} --- see for example
\cite{Delfino:Multi,Mussardo:Semiclassical} and references therein. The coupling constants $\hat\mu$ and $\hat\lambda$ are related to the parameters, $U,V_\parallel,V_\perp,t$, of the bare Hamiltonian. In particular, $\hat\lambda^2\sim (V_\parallel/2t-1)$. The term $\hat\mu\,\cos\beta\,\Phi_c$ represents the $8k_F$-Umklapp processes, and $\hat\lambda\,\cos\frac{\beta}{2}\,\Phi_c$ represents the interchain interactions.  We note that the latter term may also arise from purely 1D {\em intrachain} dimerization\cite{Essler:Quarter}. Indeed, the confining aspects of the double sine-Gordon model were first applied in Affleck's insightful study of the Heisenberg chain with frustration and dimerization\cite{Affleck:Confinement}; for related works see also\cite{Sorensen:Soliton,Augier:Soliton,Schmidt:Spectral}.

The model (\ref{dsg}), with $\beta^2=1$, has two characteristic soliton mass scales generated by the intrachain and interchain perturbations:
\begin{eqnarray}
m\equiv\Delta_{\hat\mu} & \sim & t\,\hat\mu^{1/2}\,e^{-1/\hat\mu},\\
M\equiv\Delta_{\hat\lambda} & = & {\mathcal C}\,{\hat\lambda}^{2/3},
\end{eqnarray}
where ${\mathcal C}$ is given by\cite{Zamo:Massscale,Lukyanov:Expectation}:
\begin{equation}
{\mathcal C}=\frac{2}{\sqrt\pi}\frac{\Gamma[1/6]}{\Gamma[2/3]}\left(\frac{\pi}{2}\frac{\Gamma[3/4]}{\Gamma[1/4]}\right)^{2/3}.
\end{equation}
The ratio of these scales, $M/m$, may be taken as a measure of the strength of the confinement. In particular, as we shall discuss below, the number of stable bound states decreases as $M/m$ is increased.

\begin{figure}[ht]
\centerline{\epsfxsize=10cm\epsffile[21 29 705 565]{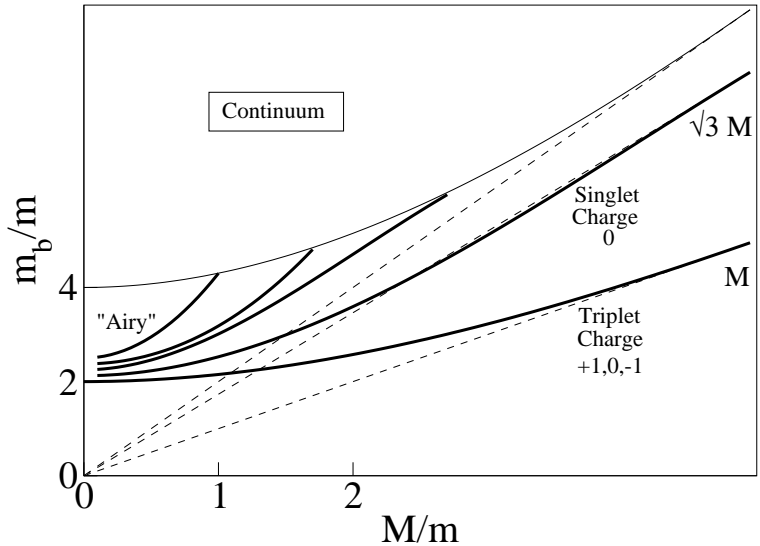}}   
\caption{Qualitative evolution of the excitation spectrum in the weakly coupled Wigner crystal.\label{FIG:QFHevol}}
\end{figure}

Perhaps the simplest regime to consider, is that with $M\gg m$. In this limit, $m$ and  $\hat\mu\rightarrow 0$ and (\ref{dsg}) yields the ``$\beta^2=1/4$'' sine-Gordon model. The spectrum of this model is known to consist of a triplet of mass $M$ and a singlet of mass $\sqrt 3 M$\cite{Lukyanov:Expectation,Dashen:Particle,Haldane:Spontaneous,Affleck:Confinement}. As such, the spectrum in the charge sector consists of four bound states. The three lightest `mesons' are degenerate in mass, and are distinguished by their charges, $\pm e,0$. In the {\em language} of the XXZ chain, these charges correspond to the projections of the auxiliary spin variable $S^z$; we emphasize that these particles are actually {\em spinless}.  Their properties are summarized in Table~\ref{mesonspec1} and in Fig.~\ref{FIG:QFHevol}. 
\begin{table}[ph]
\tbl{Bound states}
{\normalsize
\begin{tabular}{|c|c|}
\hline
Mass, $M$ & Charge, $e$ \\
\hline
$m_1=1$ & +1, 0, -1 \\
\hline
$m_2=\sqrt 3$ & 0 \\
\hline
\end{tabular}\label{mesonspec1}}
\end{table}
The relatively small number of bound states in this regime parallels the reduction to the ${\rm E}_8$ spectrum in the $h\neq 0$ Ising field theory as $m\rightarrow 0$\cite{Zamolodchikov:Integrals}. The bound states will manifest themselves as sharp peaks in the dynamical response functions. In particular, the optical conductivity (as determined by the current operator) will probe the neutral bound states. Their contributions may be extracted using the known sine-Gordon form factors. 

The regime $M\ll m$ bears a close relationship to the Ising model in a weak magnetic field. In this regime, the confining perturbation is very weak and one may (in principle) apply the Bethe--Salpeter approach of Fonseca and Zamolodchikov\cite{Fonseca:Ising}. Although we have not performed this rather technical analysis for the double sine-Gordon model, the lowest order Schr\"odinger description is applicable: 
\begin{equation}
\label{dsgschrod}
-\frac{1}{m}\frac{d^2\Psi_b(x)}{dx^2}+\lambda|x|\Psi(x)={\mathcal E}_b\Psi_b(x).\end{equation}
Solitons (whose size is much smaller than their separation) are treated as point-like particles interspersed by a region of false vacuum. This gives rise to a linear potential or `string'. The string tension $\lambda$ is related to the bare tension $\hat\lambda$ via:
\begin{equation}
\label{dsgstringt}
\lambda = 2\,\hat\lambda\,\left\langle\,\cos(\beta\,\Phi_c/2)\,\right\rangle.
\end{equation}
It may be seen that $\lambda$ has dimensions of $[{\rm mass}]^{2}$ and (\ref{dsgschrod}) is  consistent on dimensional grounds.  Note that quantum mechanical fluctuations are incorporated in the expectation value appearing in (\ref{dsgstringt}). This quantity behaves as $m^{\beta^2/2}$ and reduces $\lambda$ with respect to its classical value\cite{Affleck:Confinement} obtained as $\beta^2\rightarrow 0$. As in the weak field Ising model, the spectrum consists of a plethora of bound states, with masses given by:
\begin{equation}
\label{smallmeh}
m_b=2m+\left(\frac{\lambda^2}{m}\right)^{1/3}\,Z_b+\dots;\quad {\rm Ai}(-Z_b)=0.\end{equation}
This formula applies to both charged and neutral bound states, and the number of distinct masses is determined by the stability threshold, $m_b< 2m_1$. In this regime we therefore have many {\em generations} of `mesons' indexed by the zeroes of the Airy function --- see Fig.~\ref{FIG:QFHevol}. Each generation is four-fold degenerate and consists of a triplet with charges $\pm e,0$, and a singlet with charge $0$. This degeneracy is evidently lifted as the ratio $M/m$ increases. The evolution between these limits is depicted qualitatively in Fig.~\ref{FIG:QFHevol}. The contributions to dynamical response functions may be calculated along the lines of \S \ref{BSA}.

Having discussed the effects of confinement, let us briefly digress on the effects of a finite on-site Coulomb interaction on the {\em spin sector}. At finite $U$ there are two effects. First, there will be a Heisenberg exchange between spins on the same chain. Second, there will be a spin dependent contribution to the interchain interactions arising from the Coulomb repulsion; in the bosonization approach, the electron density operator contains a $2k_{\rm F}$ component which is spin dependent --- see for example chapter 30 of ref.\cite{Tsvelik:QFT}. In this way one may derive an interaction of the form:
\begin{equation}
V^{nm}=J\cos\left(\Phi_s^n/2\right)\,\cos\left(\Phi_s^m/2\right)\cos\left((\Phi_c^n-\Phi_c^m)/4\right).
\end{equation}
Since $\cos(\Phi_s/2)$ is the bosonized version of the staggered energy density for the Heisenberg chain, and spins at $1/4$-filling are next-nearest neighbors,it is plausible that 
\begin{equation}
 V^{nm}=J^\prime\,({\bf S}_{i}.{\bf S}_{i+2})_n({\bf S}_i.{\bf S}_{i+2})_m\cos\left((\Phi_c^n-\Phi_c^m)/4\right)
\end{equation}
is also valid when the spin exchange is much smaller than the charge bandwidth. This interaction leads to spin-Peierls ordering at low temperatures.


\section{Conclusions}
\label{CONC}
In this work we have discussed two instructive examples of confinement which occur in low-dimensions. Although such problems are generically non-integrable, field theory approaches are able to shed light on many interesting aspects. 
In closing, let us touch upon a somewhat more exotic form of confinement in a model with {\em two} scalar fields:
\begin{equation}
{\mathcal L}=\sum_{i=1}^2\frac{1}{16\pi\beta_i^2}(\partial_\mu\Phi_i)^2-g_1\cos\Phi_1+{\mathcal V}_c,
\end{equation}
where
\begin{equation}
\label{otherpert}
{\mathcal V}_c=-g_2\cos\Phi_2\cos(\Phi_1/4).
\end{equation}
First, note that $\Phi_2$ is {\em massless} in the absence of the interaction ${\mathcal V}_c$. Therefore, unlike the previous examples considered, the expectation value of the confining operator (\ref{otherpert}) vanishes in the unperturbed model. Since the mass scales for the $\Phi_1$ and $\Phi_2$ excitations are given by
\begin{equation}
m_1\sim g_1^{1/2(1-\beta_1^2)}, \quad m_2 \sim \left(g_2\,m_1^{\beta_1^2/8}\right)^{1/2(1-\beta_2^2)},
\end{equation}
the corresponding string tension $\lambda\sim m_2^2$ has a nonlinear dependence on the coupling $g_2$. Second, let us consider a pair of $\Phi_1$ solitons interpolating between $(0,2\pi)$ and $(2\pi,4\pi)$ respectively. In this background, the operator $\cos(\Phi_1/4)$ {\em changes sign},  and the interaction (\ref{otherpert}) promotes {\em massless} fluctuations of $\Phi_2$ in the region between the solitons. In this way it is possible to argue that the potential between the solitons is of the form
\begin{equation}
U(x)=\lambda |x|+\frac{Z(n_k)}{|x|}
\end{equation}
where $Z(n_k)$ is a function of the occupation numbers of the gapless modes. In this case, the $\Phi_1$ solitons are confined by a {\em double-well potential}. 
\section*{Acknowledgments}
We are extremely grateful to Fabian Essler and Robert Konik for useful discussions. This work was supported by the US DOE under Contract No DE-AC02-98 CH10886.


\end{document}